# Extended Prigogine Theorem: Method for Universal Characterization of Complex System Evolution

Sergey Kamenshchikov*

*Moscow State University of M.V. Lomonosov, Physical department,*
*Russia, Moscow, Leninskie Gory, 1/2, 119991*

**Abstract**

**Evolution of arbitrary stochastic system was considered in frame of phase transition description. Concept of Reynolds parameter of hydrodynamic motion was extended to arbitrary complex system. Basic phase parameter was expressed through power of energy, injected into system and power of energy, dissipated through internal nonlinear mechanisms. It was found out that basic phase parameter as control parameter must be delimited for two types of system - accelerator and decelerator. It was suggested to select zero state entropy on through condition of zero value for entropy production. Zero state introduces universal principle of disorder characterization. On basis of self organization $S$ – theorem we have derived relations for entropy production behavior in the vicinity stationary state of system. Advantage of these relations in comparison to classical Prigogine theorem is versatility of their application to arbitrary nonlinear systems. It was found out that extended Prigogine theorem introduces two relations for accelerator and decelerator correspondingly, which remarks their quantitative difference. At the same time classic Prigogine theorem makes possible description of linear decelerator only. For unstable motion it corresponds to strange attractor.**

TABLE OF CONTENTS

**Abbreviations**

*UPT* - Universal Prigogine Theorem; *DDF* - Density distribution function.

---

\* Corresponding author: Russia, Moscow, Krimsky val, 4, d.1, Kamenshchikov Sergey A., kamphys@gmail.com, Skype: kamenshchikov_sergey

**Introduction**

Evolution of complex system, i.e. system, containing of statistically large particles number, can be described, using stochastic state functions. This approach was firstly applied to thermodynamic systems and was extended into area of arbitrary complex systems. It was convenient to represent evolution of stochastic system as set of transitions between phase states. However, most general characterization of complex system needs introduction of *universal* measures of system disorder and system instability. It is necessary to understand connection between these two characteristics and to define their behavior in vicinity of stationary, stable states. Versatility of suggested description is impossible without proper and convenient selection of control parameter which can be used for any types of complex systems. Next three chapters are devoted to solution of these problems and goals.

**1. Stability and basic phase parameter**

Evolution of stochastic system under defined control parameter set is basic question of synergetic science. It is connected with problem of statistical description of self organization, i.e. description in terms of distribution density evolution. Stochastic system evolution may be represented as consequence of *phase states* and *phase transitions* if using terms of statistical thermodynamics.

Let's generalize these terms for an arbitrary ergodic stochastic system (ES system). If we designate $q^+$ and $q^-$ for power input and output per system volume mass, then energy balance condition can be formulated in the following way:

$$R(t) = f\left(\vec{\Pi}(t)\right) = \frac{q^+(t)}{q^-(t)} \leq 1 \qquad R_f = 1 \tag{1}$$

Here $R(t)$ is so called *basic phase parameter* and $\vec{\Pi}(t)$ is set of control parameters (characteristic vector). Quantities $q^+$ and $q^-$ correspond to power input and output per system volume mass such that $q(t) = \sum_{i=1}^{K} v_i \cdot \dot{v}_i = q^+(t) - q^-(t)$. We may use example of hydrodynamic bifurcation. Then input/output energy mechanisms are provided by flow inertial forces and by viscous dissipation correspondingly.

Reynolds number *Re* plays role of basic phase parameter in this case and is given below:

$$\text{Re}(t) = \frac{l \cdot u_0}{D(t)} = \frac{l \cdot u_0}{D(t)} \cdot \frac{dt}{dt} = \frac{\varepsilon^+}{\varepsilon^-(t)} = R(t) \tag{2}$$

Here *l* is spatial scale of system, $u_0$ is velocity of energy source (input flow) which is assumed to be constant in this example. The basic postulate of this chapter can be formulated in the following way: bifurcation necessarily corresponds to the condition: $R(t) \neq 1$, while phase condition is realized for $R = 1$ (*Hypothesis I*).

Scheme of ES – system first order phase transition then may be represented by set of chains, following below.

$$\uparrow q^+(t) \Rightarrow \uparrow R(t) \Rightarrow R(t) \succ 1 \Rightarrow \uparrow q^-(t) \Rightarrow \uparrow R(t) \Rightarrow R(t_1) = 1 \tag{3}$$

$$\downarrow q^+(t) \Rightarrow \downarrow R(t) \Rightarrow R(t) \prec 1 \Rightarrow \downarrow q^-(t) \Rightarrow \uparrow R(t) \Rightarrow R(t_1) = 1 \tag{4}$$

$$\downarrow q^-(t) \Rightarrow \uparrow R(t) \Rightarrow R(t) \succ 1 \Rightarrow \downarrow q^+(t) \Rightarrow \downarrow R(t) \Rightarrow R(t) \succ 1 \qquad (5)$$

$$\uparrow q^-(t) \Rightarrow \downarrow R(t) \Rightarrow R(t) \prec 1 \Rightarrow \uparrow q^+(t) \Rightarrow \uparrow R(t) \Rightarrow R(t_1) = 1 \qquad (6)$$

Here $\uparrow$ and $\downarrow$ show finite increase and decrease of corresponding parameter for $t_1 \succ t \succ t_0$. Initial condition of system corresponds to $R(t_0) = 1$. As it follows from set (3) - (6) positive feedback for input/output power mechanisms is compulsory condition for phase transition. Without loss of generality one can be represented in the following way: $\partial q^+(t) / \partial q^-(t) = 1$. For situation when fixed input power $q^+$ is switched to anther constant (3, 4) basic phase parameter can be represented as given expression (7).

$$R(t) = f\left(\vec{\Pi}(t)\right) = \frac{q^+(t)}{q_0^-(t_0) + q_1^-(t) + q_2^-(t) + ...} \qquad (7)$$

Here perturbation members of denominator decomposition correspond to bifurcation and deviation from one stable phase state to another one such that $R(t) = 1$ in both cases. This nonstationary process can be called phase transition if we use terminology of statistic physics.

Let's test Hypothesis I, using auxiliary entropy of Kolmogorov – Sinai $h_d$ [1]: $h_d = \langle h(\vec{x}(t))\rangle$. Here averaging in phase space is designated as $\langle\ \rangle$ and averaged quantity can be expressed as sum of positive Lyapunov factors $h_i^+$ for each dimension of generalized phase space:

$$h = \sum_{i=N}^{K} h_i^+ = \ln\left(\prod_{i=N}^{K} \sigma_i^+\right) \qquad \sigma_i^+(t) = \frac{|\delta x_i(t)|}{|\delta x_i^0|} \qquad (8)$$

Vector $\vec{x}(t)$ is characteristic phase vector of system state. Factor $\sigma_i^+$ shows distance growth $\delta x_i(t)$ in $i$ direction for two infinitely closely located points in phase space. Condition of stationary state then is equal to $h = 0$ or $\sigma_i^+ = 1$ ($i = \overline{N, K}$). Relation for specific system power is given below:

$$q(t) = \frac{\delta}{\delta t}\left(\sum_{i=1}^{K} \frac{v_i^2}{2}\right) = \sum_{i=1}^{K} v_i \cdot \dot{v}_i = q^+(t) - q^-(t) \qquad (9)$$

System stability condition leads to $h_i \leq 0$ and $\sigma_i \leq 1$ if we consider all Lyapunov factors. Given inequalities lead to expression (10) for velocity components $v_i$ $(i = \overline{1, K})$.

$$\frac{|\delta x_i(t)|}{|\delta x_i^0|} = \frac{|\delta x_i(t)/\delta t|}{|\delta x_i^0/\delta t|} = \frac{|v_i(t)|}{|v_i^0|} = \alpha(t) \qquad 0 \prec \alpha(t) \leq 1 \qquad (10)$$

Relation (10) in fact allows receiving components of acceleration $\dot{v}_i(t)$:

$$\lim_{\Delta t \to 0}\left(\frac{|v_i(t)| - |v_i^0|}{\Delta t}\right) \leq 0 \qquad \dot{v}_i(t) = \lim_{\Delta t \to 0}\left(\frac{v_i(t) - v_i^0}{\Delta t}\right) \qquad (11)$$

Indeed, consideration of specific power $q(t)$ can be reduced to two cases: a) $\delta x_i(t) \succ 0$ and $|\delta x_i(t)| = \delta x_i(t)$; b) $\delta x_i(t) \leq 0$ and $|\delta x_i(t)| = -\delta x_i(t)$. Signs of $\delta x_i(t)$ and $\delta x_i^0$ match - this condition is obligatory for definition of Lyapunov factors. Then $\alpha(t)$ doesn't depend on initial sign of coordinate shift $\delta x_i^0$.

For cases a) and b) we then receive: a) $\dot{v}_i \prec 0$ and $v_i \succ 0$; b) $\dot{v}_i \geq 0$ and $v_i \leq 0$. In both cases with use of relation (8) we receive that $q(t) \leq 0$. According to definition (7) of basic phase parameter this means that $R(t) \leq 1$ for $h_d \leq 0$. Condition of $R_f$, i.e. $R(t) = 1$ corresponds to *phase state* of system.

Let's show how realization of condition $R(t) \leq 1$ influences system stability – we test reversibility of statement, given above. Relation (1) shows that in this case $q^+(t) - q^-(t) \leq 0$ and consequently $\partial \varepsilon / \partial t \leq 0$. Here $\varepsilon$ is specific energy – energy of system mass unit. Without loss of generality this requirement leads to relation $v_i \cdot \dot{v}_i \leq 0$ $\left(i = \overline{1,K}\right)$. This condition can be reduced to two cases: a) $v_i \geq 0$ and $\dot{v}_i(t) \leq 0$; b) $v_i \prec 0$ and $\dot{v}_i(t) \succ 0$.

According to expressions (10) and (11) we have following consequences: a) $|\delta x_i(t)| \prec |\delta x_i^0|$; b) $|\delta x_i(t)| \geq |\delta x_i^0|$. Here we again use conditions of positive time delay and coincidence of $\delta x_i(t)$ and $\delta x_i^0$ signs. As it was shown in general case *R*-parameter defines necessary but not sufficient requirement of stability. *Thus use of $R(t)$ as control parameter must be delimited for two types of system: a) accelerator - $\dot{v}_i(t) \succ 0$; b) decelerator - $\dot{v}_i(t) \leq 0$. For first type of system motion stability loss and bifurcation are realized for $R(t) \prec 1$; decelerator comes to phase transition only if $R(t) \succ 1$. However for both types of motion bifurcation necessarily corresponds to the condition: $R(t) \neq 1$, while phase condition is realized for $R = 1$.* Hypothesis I has been proved.

## 2. Entropy demarcation criterion

Local self – organization $S$ – theorem, formulated by U. Klimontovich in 1983 [2] shows dependence of Lyapunov function $\Lambda_s$ from arbitrary control parameter $a(t)$ fluctuation:

$$\frac{\partial \Lambda_s}{\partial (_\Delta a)} \geq 0 \qquad \Lambda_s = \tilde{S}_2 - S_1 \qquad (12)$$

Index ~ corresponds to normalization of second state, which introduces artificial conservation of Hamilton function for both states. Normalization procedure will be particularly considered in the beginning of next chapter. In relation (12) two states of stochastic system are considered – *State 1*, corresponding to control parameter $a$ and *State 2*, which corresponds to control parameter $a = a_0 + _\Delta a$. *State 2* is more disordered state then *State 1*. In $S$ – theorem entropy is used in fact as disorder measure. However it can be used only as relative characteristic, for in $S$ - theorem entropy of regular, zero state, is not introduced, i.e. no *demarcation criterion* exists. One of possible ways for its formulation is based on bijective connection between dynamic entropy and Gibbs entropy - $h_d(S)$. Let's use separate motion of system in accelerated and decelerated stages. Then basic phase characteristic $R(t)$ can be used as the control parameter for each stage. As it was mentioned above instability takes place if $R(t) \succ 1$ for accelerator or if $R(t) \prec 1$ decelerator. Bijection $h_d(S)$ is valid in vicinity of stable phase state when $R \to 1$ and $h_d \to 0$.

Condition of energy conservation is realized in this case for $q^+(t) - q^-(t) \to 0$. Liouville theorem allows making conclusion about conservation of phase space volume and single value of entropy in vicinity of phase $i$ ($R_i = 1$). *Then zero state entropy $S_0 = 0$ can be defined from implicit condition:* $\lim_{S \to S_0} h_d(S) = 0$. *If system has continuous or discrete set of phase states $\{R_i\}$, minimum value of entropy should be selected.*

## 3. Universal Prigogine theorem

Use of demarcation criterion allows formulation of two disordered states $\tilde{S}_2$ and $S_1$, considered in $S$ – theorem: $\tilde{S}_2 \succ S_1 \succ S_0$. $S$ – theorem is realized in frame of first basic assumption that distribution functions of both states have Boltzmann form. Then index ~, applied to entropy of second state $S_2$ defines normalization, expressed by relation for Hamilton function (13):

$$\langle \tilde{H}(a + {}_\Delta a) \rangle_2 = \langle H(a + {}_\Delta a) \rangle_1 \tag{13a}$$

$$f(T) = const \cdot \exp\left(-\frac{H}{kT}\right) \tag{13b}$$

Symbol $\langle \, \rangle$ designates phase space averaging. In fact expression (13a) reveals system energy virtual conservation - normalization is achieved artificially by normalization of distribution function $f(T)$: $T \to \tilde{T}$ where $T$ is system temperature. If $\tilde{S}_2 \succ S_1$ then $\tilde{T} \prec T$, if control parameter $a(t)$ is chosen correctly. Virtual conservation means that correct comparison of system disorder is possible within such preliminary calibration of Hamiltonian. Indeed for Boltzmann form of density distribution entropy is function of system Hamiltonian $S = -\langle \ln(\rho_B(H,T)) \rangle$. Therefore comparison of two states with different distributions and values of entropy can be correct only within relation (13a), representing artificial conservation of total system energy. According to definition of dynamic entropy [1] time derivative of Lyapunov function can be represented in the following form:

$$\frac{\partial \Lambda_s}{\partial t} = \frac{\partial \tilde{S}_2}{\partial t} - \frac{\partial S_1}{\partial t} = \tilde{h}_2 - h_1 \tag{14}$$

Relation (12) can be modified for time dependent control parameter ${}_\Delta a = {}_\Delta a(t) = a(t) - a(t_0)$:

$$\frac{\partial \Lambda_s}{\partial ({}_\Delta a)} = \frac{\partial \Lambda_s}{\partial (a)} = \frac{1}{a_t} \cdot \frac{\partial \Lambda_s}{\partial t} = \frac{1}{a_t} \cdot \left(\tilde{h}_2 - h_1\right) \tag{15}$$

Consequence of $S$ – theorem is represented by relation (16). Here lower index $t$ corresponds to time derivative.

$$\frac{{}_\Delta \tilde{h}}{a_t} \geq 0 \tag{16}$$

Let's use $R(t)$ as control parameter.

Then condition (16) can be expressed in the following way in the vicinity of stationary condition $R(t) = 1$:

$$R_t \geq 0 \quad \tilde{h} - h_{st} \geq 0 \quad \dot{h} \leq 0 \qquad (16a)$$

$$R_t \leq 0 \quad \tilde{h} - h_{st} \leq 0 \quad \dot{h} \geq 0 \qquad (16b)$$

In (16a) and (16b) we have used replacement $\lim_{\Delta t \to 0} \left( \Delta h / \Delta t \right) = \dot{h}$ after dividing both sides of inequality (16) by $\Delta t$. Transition $2 \to 1$ to stationary stable state is considered. According to conclusion of Chapter 1 relation (16a) corresponds to decelerator type of motion, when stability loss is caused by increase of basic phase parameter $R(t): R_t \geq 0$. (16b) describes the case of accelerated motion - $\dot{v}_i(t) \geq 0$.

Relations (16a) and (16b) give content of generalized Prigogine theorem [5] for $R(t)$ control systems. We used hypothesis that value of control parameter for stable state matches control parameter of stationary motion. I.e. we assume that relaxation of stable system to stationary state finally occurs if control parameter stays fixed.

Unlike Prigogine theorem we have achieved statement with general area of application. Indeed, relations (16a), (16b) are valid for systems with arbitrary class of linearity.

Classic Prigogine theorem defines entropy production in vicinity of stationary state. However, conclusion of the theorem is based on linear approximation of series expansion for control parameter component:

$$\dot{a}_\alpha = l_{\alpha,\beta} \cdot \left( \frac{\partial F}{\partial a_\beta} \right) + k_{\alpha\beta} \left( \frac{\partial F}{\partial a_\beta} \right)^2 + \ldots \qquad (17)$$

Here $F$ is thermodynamic free energy and $l_{\alpha,\beta}$ are constant kinetic factors of Onsanger relations [5]. In frame of linear expression relation (16a) is achieved – classical Prigogine theorem describes linear decelerated type of system. *That is why advantage of (16a) and (16b) relations in comparison to classical Prigogine theorem is versatility of its application to arbitrary systems controlled by basic phase parameter $R(t)$.*

Let's list basic assumptions of Nonlinear Prigogine Theorem, which were used above:
- $\Delta t \succ 0$. This condition is natural for physical description of system evolution;
- Density distribution function (DDF) has Boltzmann form $f = f_B = const \cdot \exp\left( -\frac{H}{kT} \right)$. Here $H$ is Hamilton function of system and $T$ is its temperature. *DDF* is valid for systems with independent particles trajectories in phase space. This condition is valid for chaos state [4], when $h_d \succ 0$ and phase space resolution is finite - $\Delta \Gamma \succ 0$. Here $\Delta \Gamma$ is element of phase space;
- $\left\langle \tilde{H}(a + \Delta a) \right\rangle_2 = \left\langle H(a + \Delta a) \right\rangle_1$. Condition of virtual energy conservation is reached by normalization of system temperature - $T_2 \to \tilde{T}_2$.

**Conclusion**

Evolution of arbitrary stochastic system was considered in frame of phase transition description. Concept of Reynolds parameter of hydrodynamic motion was extended to arbitrary complex system and basic phase parameter $R(t) = \dfrac{q^+(t)}{q_0^-(t_0) + q_1^-(t) + q_2^-(t) + ...}$ was introduced. We came to conclusion that use of $R(t)$ as control parameter must be delimited for two types of system: a) accelerator - $\dot{v}_i(t) \succ 0$; b) decelerator - $\dot{v}_i(t) \leq 0$. For first type of system motion stability loss and bifurcation are realized for $R(t) \prec 1$; decelerator comes to phase transition only if $R(t) \succ 1$. However for both types of motion bifurcation necessarily corresponds to the condition: $R(t) \neq 1$, while phase condition is realized for $R = 1$.

Entropy as disorder measure is commonly applied for characterization of complex system state. However arbitrariness of additive constant leads to absence of universal approach to disorder characterization. In the current work it was suggested to define zero state entropy $S_0 = 0$ from implicit condition: $\lim_{S \to S_0} h_d(S) = 0$. For continuous or discrete set of phase states $\{R_i\}$, minimum value of entropy has to be selected.

On basis of self organization $S$ – theorem we have derived relations for entropy production behavior in the vicinity stationary state of system. That is why advantage of these relations in comparison to classical Prigogine theorem is versatility of their application to arbitrary systems controlled by basic phase parameter $R(t)$. Classical Prigogine theorem describes only linear decelerated type of system. Extended Prigogine theorem defines two relations for decelerated motion ($\dot{v}_i(t) \prec 0$) and accelerated one ($\dot{v}_i(t) \geq 0$). First type corresponds to $\tilde{h} \geq h_{st}$ connection – minimum entropy production is achieved in stationary, stable phase state. Second type of motion is described by opposite inequality: $\tilde{h} \leq h_{st}$. It could be useful to remark qualitative and quantitative difference in accelerated and decelerated motion. Both of these types could be unstable. But in one case restoring force ($\dot{v}_i(t) \prec 0$) corresponds to attraction into certain phase space. In another case we have repulsion as determinative mechanism. If we consider unstable trajectories first type of motion corresponds to strange attractor and second one to repeller. Extended Prigogine theorem gives opposite scenarios of evolution in vicinity of phase state for these two types of motion.